# Controlled Nucleation and Growth of CdS Nanoparticles in a Polymer Matrix

*Tiziana Di Luccio**

ENEA, Centro Ricerche Portici, Via Vecchio Macello, I-80055 Portici (NA), Italy

*Anna Maria Laera, Leander Tapfer*

ENEA, Centro Ricerche Brindisi, SS7 Appia Km 706, I-72100 Brindisi, Italy

*Susanne Kempter, Robert Kraus, Bert Nickel*

Department für Physik and CeNS, Ludwig-Maximilians-Universität, Geschwister-Scholl-Platz 1, D-80539 München, Germany

* Corresponding author. E-mail: tiziana.diluccio@portici.enea.it

Telephone: +39 081 7723244  Fax: +39 081 7723344

**Abstract**

In-situ synchrotron X-ray Diffraction (XRD) was used to monitor the thermal decomposition (thermolysis) of Cd thiolates precursors embedded in a polymer matrix and the nucleation of CdS nanoparticles. A thiolate precursor/polymer solid foil was heated to 300 °C in the X-ray diffraction setup of beamline W1.1 at Hasylab and the diffraction curves were recorded every 10 °C. At temperatures above 240 °C, the precursor decomposition is complete and CdS nanoparticles grow within the polymer matrix forming a nanocomposite with interesting optical properties. The nanoparticle structural properties (size and crystal structure) depend on the annealing temperature. Transmission Electron Microscopy (TEM) and Photoluminescence (PL) analyses were used to characterize the nanoparticles. A possible mechanism driving the structural transformation of the precursor is inferred from the diffraction features arising at the different temperatures.

**Introduction**

The controlled growth of wide band gap II-VI semiconductors with nanometric dimensions has become an important issue in material science research. The optical and electronic properties of nanoparticles are unique[1] and open possibilities of applications in several fields, such as antireflective coatings,[2] emitting devices in the yellow/red spectral region,[3] biological labels[4] or bioelectronics.[5,6] Among the different II/VI semiconductors, CdS has a bulk band gap $E_G$ = 2.45 eV. A shift $\Delta E_G$ = 0.8 eV with respect to the bulk value is observed in 3 nm nanoparticles, that is $E_G$ = 3.25 eV;[1] such a large value of $E_G$ allows light emission between blue and red wavelengths.

To synthesize nanoparticles, methods using inverse micelles,[7] or the reaction of organometallic reagents dissolved in coordinating solvents,[8–10] or aqueous solution synthesis[11–14] have been developed. In the last method, thiols are often used as stabilizing agents for CdS,[11,13] CdSe,[12] and CdTe[14] nanoparticles. An alternative method (not involving hazardous reagents) is based on the thermolysis of metal thiolate molecules. In a thermolysis process, a metal thiolate powder is heated beyond its melting point; the molecules rearrange and finally decompose. The decomposition of metal thiolate precursors into crystalline metal and metal sulfides has been reported in the literature to produce nanoparticles of CdS,[15] ZnS,[16] PbS and HgS,[17] and Au.[18] Typically, annealing temperatures of 200 °C and annealing times of 1–2 hours are required to produce spherical nanocrystals, while annealing temperatures of 140–190 °C favor the formation of nanorods.[19]

During the last years, increasing attention has been devoted to nanocomposites, for example a transparent polymer matrix with inorganic nanoparticles embedded for their many interesting applications in optoelectronics.[3,20,21] To fabricate such a nanocomposite, the nucleation of nanoparticles directly in the polymer matrix can be useful if it is possible to control the nanoparticle dispersion and size within the polymer matrix, thus suppressing segregation.

In the present work we report on the nucleation and growth of cadmium sulfide (CdS) nanoparticles in a polymer matrix using thermolysis of metal thiolates. We follow the particle formation on the molecular scale using a combination of structural, optical and chemical probes. The main aspects of the structural properties of the CdS nanoparticles obtained by this method were reported in a previous work where ex-situ small and wide angle X-ray scattering (SAXS and WAXS, respectively) were performed on samples annealed at different temperatures.[22] The present work consists of an in-situ study devoted to understand what mechanism leads to the transformation of the thiolates into the nanoparticles by means of synchrotron X-ray Diffraction (XRD). Topas (TP) and polystyrene (PS) are used as the polymer matrix since these polymers are optically transparent and chemically inert. As precursors, we synthesized Cd thiolate molecules $Cd(SR)_2$ with *n*-alkyl thiols of different chain length depending on the number *n* of carbon atoms: $C_nH_{2n+1}SH$ (*n*= 3, 5, 12, 18). The precursors were dispersed in a Toluene-polymer solution. Then, the solution was dried, and the plastic pieces obtained in this way were annealed in vacuum to temperatures up to 300 °C. The nomenclature we use for the samples is C3/TP, C5/TP, C12/TP, and C18/TP, thus indicating the chain length of the corresponding precursor and the matrix material. The paper is organized as follows. First, we discuss the distribution and structure of the precursor molecules in the polymer matrix before the thermolysis process. This information is inferred from X-ray measurements. Second, TEM and photoluminescence spectra (demonstrating the formation of CdS nanoparticles of controlled size using thermolysis) are presented. Finally, the detailed transformation of the precursor molecules into CdS nanoparticles during the thermolysis process is followed by means of in-situ synchrotron XRD for the example of C12/TP. In the discussion we rationalize the molecular mechanism of the nanoparticle formation based on these measurements.

**Experimental Section**

**Preparation of CdS nanoparticles in topas.** The CdS/polymer nanocomposites were prepared by thermolysis of cadmium thiolate precursors $Cd(SR)_2$ dispersed in polystyrene (PS) [23,24] and topas (TP).[22] The data presented in this work focus on CdS nanoparticles dispersed in TP;

however similar results have been obtained using PS. TP is a cycloolefin copolymer consisting of ethylene and norbonene chains and it is distributed by Ticona. Due to its excellent transmittance, combined with low autofluorescence and extremely low water absorption, TP is suited for biological and optical applications. TP was provided by Ibidi, München.

Four $Cd(SR)_2$ precursor molecules (R = $C_{18}H_{37}$, $C_{12}H_{25}$, $C_5H_{11}$, $C_3H_7$) were synthesized using (*n*-alkyl) thiol molecules and $Cd(NO_3)_2$ as the starting material.[23] After synthesis, the precursor molecules (powders) were dispersed in a 10 mL polymer solution and a film was obtained by casting. The precursor/polymer weight ratio was fixed at 28% (w/w ratio) for all the samples. The solvents used were chloroform for PS and toluene for TP solutions, respectively. The final $Cd(SR)_2$/polymer foil appears homogeneous and white opalescent. Thermal annealing of the foil above 200 °C (in vacuum) leads to the formation of nanosized CdS crystals in the polymeric matrix.[22] Once the nanoparticles form, the foil changes color from white to lemon and dark orange, as the nanoparticle size increases with increasing temperature.

**Characterization techniques.** The structural properties were investigated by X-ray scattering. Ex-situ wide angle X-ray scattering measurements (WAXS) were acquired in-house. The SAXS/WAXS instrument at LMU is equipped with a Mo microfocus X-ray source purchased by Molymax Osmic ($\lambda$ = 0.71 Å, flux of about $10^6$ photons/s). This wavelength ensures low absorption; furthermore the whole system is under vacuum to reduce air scattering. The scattered signal is recorded by an image plate with a pixel size of 50 μm. The duration of each measurement has to be at least one hour in order to obtain a good signal-to-noise ratio. In-situ X-ray diffraction experiments were carried out using a fixed energy of 10.5 keV at the beamline W1.1 at Hasylab, Hamburg. In this case, the samples consisted of a precursor/polymeric foil of an area of about 4 $cm^2$ and a thickness of 1 mm. The sample was positioned in vacuum on a copper support heated by resistors. The setup reaches temperatures up to 300 °C. The temperature was stabilized by a Lakeshore temperature controller. Data were acquired in temperature steps of about 10 °C by scanning the detector while the angle of the incident X-ray beam was fixed at 2°. Each scan took about 12 min. Additional transmission electron microscopy (TEM) measurements were performed using a 100 keV JEOL system. The samples for TEM were prepared by dissolving 7 mg of annealed precursor/polymeric foil (containing the nanoparticles) in 0.2 to 0.6 mL of toluene. This solution was drop cast on Cu-grids of 3.05 mm diameter and 300 meshes. The optical properties of annealed samples were accessed by room-temperature photoluminescence (PL) measurements performed by a Hamamatsu Streak Camera System C5680, equipped with a frequency tripled Nd:YAG semiconductor pumped solid-state laser (exciting wavelength 355 nm). Gas chromatography/mass spectrometry (GC/MS) was employed to test for side products of the thermolysis process.

**Results and Discussion**

**Structure of Cd(SR)$_2$ Precursors.** For long chain thiols ($n = 12, 18$), the chain–chain interaction of the Cd thiolate precursors favors a lamellae structure of the unit cell as drawn in (Figure 1a). Three molecular configurations (cases i, ii, iii) can induce a multilayer with identical periodicity $D$ along the chain direction ($y$). In configuration (i), the molecule is formed by one Cd atom bound to two S atoms and it is folded back toward one side. In configuration (ii) the Cd atom is bound to two S atoms along the $y$-direction, while in (iii) they form an armchair configuration. Along $y$-direction such a periodic structure gives rise to well-pronounced Bragg peaks ("SLi") as observed in the WAXS curves for the long-chain precursors C12 ($n = 12$) and C18 ($n = 18$) (Figure 1b). For shorter chain thiols ($n = 3, 5$), the symmetry of the unit cell should be mainly determined by coordination chemistry of Cd and S. In the last case, a (tetragonal) unit cell should form such that two Cd atoms tetrahedrically coordinate to four S atoms, each of them bound to a hydrocarbon chain.[25,26] The structure corresponding to the four precursors ($n = 3, 5, 12, 18$) dispersed in TP as measured by WAXS is summarized in [Figure (1b)]. The TP has a large amorphous reflection at $q = 12.5$ nm$^{-1}$; the rest of the peaks is due to the precursor structure. For the precursors with lamellae structure (C12 and C18), the data show that as the chain length increases, the period $D$ of the multilayer increases as well (from 35 Å to 50 Å) and thus the separation of the Bragg peaks decreases, respectively. We compared these values of the chain periodicity with the values $D_c$ calculated by projecting the bond lengths[27] of the carbon atoms along the linear direction of the molecule and adding up the atomic diameters of S and Cd (see Table 1). The short molecule precursors (C3 and C5) do not show the lamellae structure, since no multilayer peaks are observed in Figure 1b. Thus, we estimate the interplanar distance from the first diffraction peak and compare it with the calculated chain length $D_c$. We find better agreement for long chains (C12 and C18) than for short chains (C3 and C5) as expected. An additional Bragg peak is observed around $q = 15$ nm$^{-1}$, present both in C12 and in C18. This peak is probably related to the Cd−Cd distance in the x-direction of the molecular structure (Figure 1a). This distance measures 4.16 Å and 4.22 Å for C12 and C18, respectively. These values are slightly larger than the Cd-Cd distance (3.32 Å) reported by Bayón et al.[25] for shorter molecules of cadmium thiolate with a tetragonal unit cell. We are able to observe the peak at $q = 15$ nm$^{-1}$ only in the lamellar structure. The molecules of C3 show a much more complicated structure as seen from the WAXS curve, and C5 seems to have some intermediate structure among the short and the long molecules.

**Physical properties of CdS Nanoparticles in Topas. Case of C12/TP.** *TEM.* After the thermal annealing the precursor molecules transform into CdS nanoparticles. The local structure of the nanoparticles embedded in the topas matrix is shown in the TEM images in Figure 2. The TEM images refer to a C12/TP sample annealed at 276 °C. The image in Figure 2a was recorded at 40 K magnification, and the scale (70 nm) is indicated in the figure. The selected area is then zoomed in Figure 2b at 100 K magnification. The different contrast reveals the CdS nanoparticles with respect to the topas matrix, as indicated by red circles in Figure 2b. The images suggest that the nanoparticles have a spherical shape. Their typical diameter at this temperature is estimated to be c.a. 3 nm. The TEM on samples annealed at 300 °C shows that the nanoparticles are aggregated and in many cases it is not possible to estimate a characteristic size.

**Photoluminescence.** The photoluminescence (PL) properties of four samples of C12/TP annealed at temperatures of 275, 245, 240, and 230 °C are shown in Figure 3. The inset contains a picture of the samples under UV lamp excitation ($\lambda$ = 350 nm), where we observe a gradual change from orange to blue emission when the temperature is decreased. Correspondingly, the mean diameter value of the nanoparticles diminishes from 4.4 nm to about 1 nm. The PL curves have been normalized to take into account different exposure times and different detector amplifications. The measurements show a broad peak that blue-shifts with the annealing temperature from about 610 to 480 nm. If the annealing temperature is increased above 280 °C, then the PL signal becomes very weak and reduces to background. Broad emissions in such a wavelength range are usually attributed to charge carrier recombination in trap states.[7,28−31] In some cases, trap emission has been related to low crystallinity,[29] in other cases to sulfur excess or other defects at the interface of the nanocrystals.[7,30] However, trap and band-edge emissions are often observed together in CdS nanocrystals.[31,32] Increasing the amount of surface defects of the nanoparticles can enhance enormously the trap state emission with respect to the band-edge recombination mechanism[32] that can be completely suppressed, as is likely to happen in our samples. The blue-shift of the PL peak in the direction of smaller nanoparticle size (in our samples decreasing with lower annealing temperatures) is also confirmed by similar results in ZnS[33] and CdS[34] nanoparticles. We are currently investigating the PL properties of our samples with further experiments (luminescence lifetime) that will be the object of a separate paper.

**X-ray diffraction.** We discuss now the X-ray diffraction data as a function of temperature for the C12/TP sample to infer the nucleation and growth mechanism of CdS nanoparticles from the thermolysis of C12. Figure 4 shows the temperature dependence of the in-situ synchrotron XRD data of the C12/TP sample as the function of annealing. In the temperature range of 100−200 °C the interesting part of the spectrum is the low $q$-region $(0 < q < 20)$ nm$^{-1}$ (Figure 4a). At 100 °C the peaks labeled by SL$i$ ($i$ =1–5) are identified with the lamellae precursor structure reported in (Figure 1a). The broad peak at $q$ = 12.5 nm$^{-1}$ is the TP amorphous peak. The X-ray diffraction curve at T =

100 °C (Figure 4a) shows regularly spaced, well-pronounced peaks indicated by SLi due to the multilayer lamellae structure of the C12 molecule. The lamellae structure of the C12 changes by the effect of the heating, as seen from the decrease of the SL peak intensity when the temperature is raised. On the contrary, the position of the TP peak remains constant, showing that the TP structure does not change with the temperature. The SLi Bragg reflections correspond to a unit cell of about 35 Å. An estimation of the quantity of precursor needed for the formation of the nanoparticles can be obtained from the width of the SL peaks of (Figure 4a). For example, from SL1, a domain size of 24 nm is calculated implying that only seven molecular layers of C12 may be enough to produce the CdS nanoparticles. As the temperature is increased, the SL peak intensity decreases and finally vanishes at 160 °C; at this point the lamellae structure is gone. At the same time, new peaks Pi ($i$ = 1–4) appear between 100 and 240 °C. The new peaks are transitory: P1 appears at 100 °C and disappears above 160 °C, P2 is present from 150 to 240 °C, P3 and P4 from 160 to 240 °C. Such transitory peaks arise from intermediate states of the decomposition of the precursor, since they disappear at 240 °C where the decomposition is completed, according to thermogravimetry and differential thermal analyses previously performed on the Cd thiolate powder.[23] The peaks P2, P3, and P4 are compatible with the structure factors of (11), (02) and (22) reflections of a hexagonal lattice with a centered rectangular unit cell of lattice parameters $a$ = 4.45 nm and $b$ = 2.55 nm. Our interpretation is that between 160 °C and 240 °C the precursor hydrocarbon chains are flexible or partially decomposed by the effect of the heating causing the cleaving of the lamellae structure. It has to be noted that another peak ("P") at 200 °C is very well defined and its position corresponds to a distance in real space of 26 Å, about ½ of the C12 molecule length. At the same temperature, no diffraction signal from CdS nanocrystals is present. The temperature dependence between 200 and 300 °C is shown in parts b and c of Figure 4. In the low $q$- region (Figure 4b), the peak P shifts toward lower $q$ (larger distances) as the temperature is increased. The position of P is related to the mean distance $d$ among the nanoparticles through the relation $d = 2\pi/q$, where $q$ is the scattering vector corresponding to P. We find that the value of $d$ is of the order of magnitude of the nanoparticle diameter, at least for $T < 280\ °C$ where we are able to observe the peak. Starting from 280 °C the distance among the particles is too large and it is beyond our resolution. From the full width at half maximum (fwhm) of the peak P, the domain size over which the nanoparticles arrange within the polymeric matrix can be evaluated. For example, for the diffraction curve at $T = 250\ °C$, $\Delta q = 0.7\ nm^{-1}$ implies that the nanoparticles are regularly distributed in a domain of about 9 nm.

The crystalline CdS nanoparticle growth is detectable by X-ray at 230 °C when the (111) reflection from cubic CdS appears in the diffraction curve (Figure 4c). The high $q$- region shows that the formation of CdS nanocrystals starts at 230 °C (Figure 4c); at 250 °C, three CdS peaks arise at $q$- values of 18.88, 30.44, and 35.55 $nm^{-1}$ matching with the zinc blende (111), (220), and (311) reflections of CdS. Above 250 °C, not only do the CdS peaks become narrower but also additional peaks are detected. The positions of the new peaks correspond to a wurtzite structure of CdS.[35] The nanoparticle diameter increases from about 1 to 8 nm with the annealing temperature, as determined by the simulation of the X-ray diffraction data in (Figure 5). We simulated the data by considering the nanoparticles as spherical crystallites with mixed zinc blende and wurtzite phases as often reported in the literature.[8,35] We report the results of the simulations for the diffraction curves at $T = 240\ °C$ and $T = 300\ °C$ in parts a and b of (Figure 5a) and (Figure 5b), respectively, where the suffixes $zb$ and $w$ were used to indicate the reflections of zinc blende and wurtzite phases. Where the reflections from both phases are superimposed, the relative Miller indexes are reported in the same position (e.g. $(002)_w$ and $(111)_{zb}$). The model we used for the simulation of the experimental diffraction patterns is based on the theoretical approach of Warren that considers crystals that are small in size (geometrical extension < extinction length) and spherical in shape.[36] The scattered X-ray intensity of a spherical shaped nanoparticle of diameter $D$ can be written as

$$I^{zb,w}(q) = \sum_{hkl} m_{hkl} \cdot L_{hkl} \cdot \left|F_{hkl}^{zb,w}\right|^2 \cdot \frac{\pi \cdot D^4}{2 \cdot s(q)^2} \cdot \left[1 - \frac{2}{s(q)} \cdot \sin(s(q)) + \left[\frac{2}{s(q)} \cdot \sin(\frac{s(q)}{2})\right]^2\right] \quad (1)$$

with

$$s(q) = 2 \cdot \pi \cdot q \cdot D$$

where $q$ is the scattering vector, $L_{hkl}$ is the Lorentz-Polarization factor, and $m_{hkl}$ and $F_{hkl}$ are the multiplicity and structure factor of the *hkl* reflection, respectively. For a composite material consisting of nanoparticles of both wurtzite and zinc blende crystal phases, the total scattered X-ray intensity $I_{total}(q)$ is given by the sum

$$I_{total}(q) = x \cdot I^{zb}(q) + (1-x) \cdot I^{w}(q) \qquad (2)$$

where $x$ is the density ratio between the zinc blende and wurtzite phases. In all our calculations we used $x = 0.5$. Moreover, the calculation is complicated by the contribution of the polymer. We chose to subtract a parabolic background. The diffraction patterns were simulated by using eq2 and employing a trail and error algorithm to achieve the best fit of the experimental data. As an example, in Figure 5 we show the results relative to the annealing temperatures 240 and 300 °C. The nanoparticle diameter was evaluated to be Ø = 1.8 nm ($T$ = 240 °C) and Ø = 8 nm ($T$ = 300 °C). In general, at $T \leq 250$ °C the nanoparticles seem to have zinc blende phase, even though the wurtzite and zinc blende components are not distinguishable because of the reduced dimensions of the crystallites (Figure 5a). The reflections from both phases are more and more separated at $T > 250$ °C and can be uniquely indexed at $T = 300$ °C (Figure 5b). The details of the simulations for the diffraction curves at 250 and 280 °C are reported in the Supporting Information (Figure S1).

The XRD and PL data as a function of the annealing temperature are summarized together in Figure 6. On the left axis we plot the positions of the PL maximum from Figure 3, and on the right axis the crystallite size (diameter) of the CdS nanoparticles evaluated from the simulations of the diffraction measurements is plotted (Figure 5, Figure S1); the straight lines connecting the data points are drawn as an eye guide only.

The mechanism of the precursor decomposition and the CdS nanoparticles nucleation is summarized in the cartoon of Figure 7. The initial structure (room temperature up to 100 °C) is formed by lamellae that repeat with a period $D = 35$ Å given by the sum of two adjacent $Cd(SR)_2$ molecules of C12 dispersed in the TP matrix [Figure (7a)]. This structure gives rise to the SL peaks in (Figure 4a). At intermediate temperature ($T = 160–240$ °C), the Cd atoms form an hexagonal lattice with lattice parameters $a$ and $b$ as above (Figure 7b). The side of the hexagon ($b = 2.55$ nm) is comparable with the 2.6 nm distance measured at 200 °C from the peak P (Figure 4a). In the final state (above 240 °C), the thermal decomposition of the precursor is completed, with the formation of CdS atoms and their aggregation in nanoparticles dispersed in the polymer matrix (Figure 7c).

**Gas chromatography/mass spectrometry.** Additional information about the chemical reaction of $Cd(SR)_2$ under heating was obtained by gas chromatography/mass spectrometry (GC/MS). The sample consisted of few milligrams of C12 powder annealed at 200 °C under nitrogen flux. At this temperature in nitrogen atmosphere some CdS nanoparticles are already formed and it was possible to isolate the side products of the reaction by filtering the CdS particles from the methanol solution. It is worthwhile to be reminded here that we call C12 the molecule with $R = C_{12}H_{25}$. From the GC/MS analyses we observed that the main product is the sulfide $SR_2$: MS $m/e$ (%) 370 ($M^+$, 10.5), 257 ($M^+$, 3.2), 215 ($M^+$, 4.7) and 201 ($M^+$, 100). Traces of the disulfide $(SR)_2$ are also detected: MS $m/e$ (%) 402 ($M^+$, 100), 201 ($M^+$, 17.6), 215 ($M^+$, 5.8), and 234 ($M^+$, 5.3). When the annealing was carried out at $T = 300$ °C in vacuum, we were not able to isolate the side products. In this case, we compared thin-layer chromatography of solutions of C12 annealed at different temperatures: 200, 240, and 300 °C. While a well defined spot of sulfide was observed in the sample annealed at 200 °C in nitrogen, the same spot was very weak in the sample annealed at 240 °C and disappeared for the sample annealed at 300 °C in vacuum. Therefore this suggests that above 240 °C the sulfide decomposes in more volatile products[16] that are removed by the vacuum pumping and consequently are not detected by XRD. The only products of the thermolysis are the CdS nanoparticles embedded in the polymer matrix.

**CdS nanoparticles from other thiolate precursors.** Beyond C12/TP, we studied the dynamic of the thermolysis process of the other precursors (C18, C5, and C3) dispersed in TP. The in-situ diffraction data showed that the behavior of C18/TP is very similar to C12/TP (Supporting Information, Figure S3). In particular, the intermediate peaks appear and disappear at the same temperatures. Therefore, the growth mechanism of CdS nanoparticles and their structural evolution should be the same in C18 as in C12. On the contrary, the thermolysis process shows some differences in C5 and C3 (Support Information, Figures S4 and S5). First of all, the thermal treatment does not cause the melting of the polymeric foil at the same temperature as in C12 ($T = 120$ °C) but at higher temperatures, the precursor structure remains the same from room temperature to about 200 °C. At $T = 250$ °C, the precursor is not completely decomposed, but the nanoparticles are already well defined and they form directly in the cubic−hexagonal mixed phase. Moreover, no packing peak P characterizes the nanoparticle separation in C3, as the one observed for C5, C12, and C18. Since the nucleation process was more difficult to control in C3/TP, in this sample, we investigated the dependence of the structural evolution on the annealing time (Supporting Information, Figure S6). We fixed the temperature at four different values around 240 °C (temperature of the nanoparticles nucleation). We repeated the scan twice at 220 and 230 °C, four times at 240 °C and once at 250 °C, waiting 5 minutes between each measurement. The time interval between the first and the fourth scan at 240 °C was 56 min. and at the final structure is comparable with the one at 250 °C obtained by increasing the temperature in one step. The structure of C3/TP changes from the precursor to cubic−hexagonal CdS without passing through the pure cubic phase. The precursor decomposition is completed at a temperature about 10 °C higher than that in C12, because the molecule is more stable due to the short length of the chain. In general, the process is more controlled with C12 and C18 than with C3 and C5. To have direct proof of the kinetic of the process, we annealed at the same temperature (232 °C) four precursor/TP foils and we observed a different fluorescence under UV illumination, from orange (C3) to yellowish (C5) green (C12) and white (C18). This result indicates that at 232 °C the nanoparticles are larger for shorter chains. For such short molecules, the nanoparticle formation is faster and less under control, which represents a drawback for their application.

**Conclusions**

In conclusion, the chemical and structural transformation of $Cd(SR)_2$ molecules into luminescent CdS nanoparticles dispersed in polymer matrices was followed mainly by in-situ synchrotron X-ray diffraction. The size and the mutual distance among the nanoparticles can be easily controlled in long thiolate precursors. Supporting analyses (TEM, PL, and GC/MS) contributed to characterize the formation process and the nanoparticle properties. These experiments showed that the nucleation of CdS nanoparticles is driven by the transformation of the precursor molecules. Such transformation is induced by thermal annealing of the precursor/polymeric foil and does not depend on the type of polymer used. Further investigations are in progress to clarify the optical properties of photoluminescence that indicate charge carriers recombination in trap states.

**Aknowledgement**

This work was supported by (a) the Ministero dell'Istruzione, dell'Università e della Ricerca (FISR project *Synthesis of Nanophases and Nanostructured Materials*); (b) European Community - Research Infrastructure Action under the FP6 program *Structuring the European Research Area* through the Integrated Infrastructure Initiative *Integrating Activity on Synchrotron and Free Electron Laser Science*; (c) a contribution of Consiglio Nazionale delle Ricerche (CNR) and Deutsche Forschungsgemeinschaft (DFG) within the scientific cooperation between CNR and DFG. We thank Marzia Pentimalli and Francesco Antolini for constant and helpful discussions.

**Supporting Information Description**

Details of the simulations of the XRD curves at $T$ = 250 and 300 °C (Figure S1), a TEM image of C12/TP annealed at 300 °C (Figure S2), the in-situ temperature dependence XRD curves for the other precursors/polymer foils, C18/TP (Figure S3), C5/TP (Figure S4) and C3/TP (Figure S5), and the in-situ annealing time dependence XRD data relative to C3/TP (Figure S6). This material is available free of charge via the Internet at http://pubs.acs.org.

.

**Tables**

**Table 1. Chemical Formula and Structural Data Relative to the Four Thiolate Precursors Used**

| $Cd(SR)_2$ precursor | R | $D^{\,a}$ (Å) | $D_c^{\,b}$ (Å) |
|---|---|---|---|
| C3 | $C_3H_7$ | 8.84 | 10.10 |
| C5 | $C_5H_{11}$ | 14.54 | 15.71 |
| C12 | $C_{12}H_{25}$ | 35.03 | 35.56 |
| C18 | $C_{18}H_{37}$ | 50.67 | 50.46 |

$^a$ $D$ represents the period of the lamellae multilayer structure for C12 and C18, while it is the inter-planar distance corresponding to the first diffraction peak for C3 and C5. $D$ is obtained from WAXS measurements of Figure 1b. $^b$ $D_c$ is the molecule length as calculated from the bond lengths (Figure 1a).

**Figure captions**

**Figure 1.** (a) Three similar configurations (i, ii, iii) of the lamellae structure of the long-chain (e.g., C12) molecules. The lamellae form a multilayer of periodicity $D$ in the $y$-direction. (b) WAXS data of all the precursors dispersed in TP. The measurements refer to the polymeric foils before the annealing. The broad TP peak is at $q = 12.5$ nm$^{-1}$, all of the other reflections are due to the precursor structure.

**Figure 2.** Transmission electron microscopy (TEM) images of C12/TP after annealing at $T = 276$ °C. (a) The dark spots are the CdS nanoparticles dispersed in the topas matrix; and (b) zoom of the region included in the square box. Some nanoparticles are highlighted in red circles.

**Figure 3.** Photoluminescence (PL) emission from C12/TP annealed at temperatures $T = 275, 245, 240$ and $230$ °C (excitation $\lambda = 355$ nm). A blue shift is observed when the temperature is decreased. In the inset a picture of the different emission colors from the samples illuminated under UV lamp is shown.

**Figure 4.** Synchrotron X-ray diffraction curves of C12/TP as a function of annealing temperature. (a) $T = (100–200)$ °C. The C12 precursor peaks are indicated by SLi. The peaks Pi ($i = 1, 4$) are transitory peaks induced by the thermolysis of the C12. **P** is related to the mean distance among the growing CdS nanoparticles, and TP is the topas peak. (b) $T = (230–300)$ °C, $q = (1–8)$ nm$^{-1}$ and (c) $T = (230–300)$ °C, $q = (10–40)$ nm$^{-1}$. The $q$- range is plotted in two parts, the high $q$ regime, where the crystalline reflections from CdS are measured, and the small $q$ regime, where the decomposition of the precursor is followed.

**Figure 5.** Theoretical simulations of the in-situ synchrotron XRD data of C12/TP at annealing temperatures $T = 240$ °C (a) and $300$ °C (b), by considering spherical CdS nanoparticles with zinc blende (*zb*) and wurtzite (*w*) phases of diameter Ø = 1.8 nm (a), and 8.0 nm (b). In both (a) and (b): (i) the dots represent the experimental data points, the green curve is calculated for nanoparticles with *zb* phase, the blue curve for nanoparticles with *w* phase, and the red curve is obtained by considering a 50% *zb* and 50% *w* mixed phase; (ii) the peaks are labeled by the relative Miller indexes of the *zb* and *w* phases.

**Figure 6.** PL maximum and crystal size of the CdS nanoparticles as a function of the annealing temperature.

**Figure 7.** Schematic view of the chemical transformation of the thiolate precursors and the nucleation and growth of the CdS nanoparticles from the precursor thermolysis: (a) lamellae structure of the precursor from room temperature up to 100 °C; (b) hexagonal structure due to the flexibility of the alkyl chains in the temperature range between 160 and 240 °C; (c) final state (above 240 °C) consisting of CdS nanoparticles only. The organic component has completely decomposed. The crystalline structure of the nanoparticles has been simplified in the sketch.

**Figure 1**

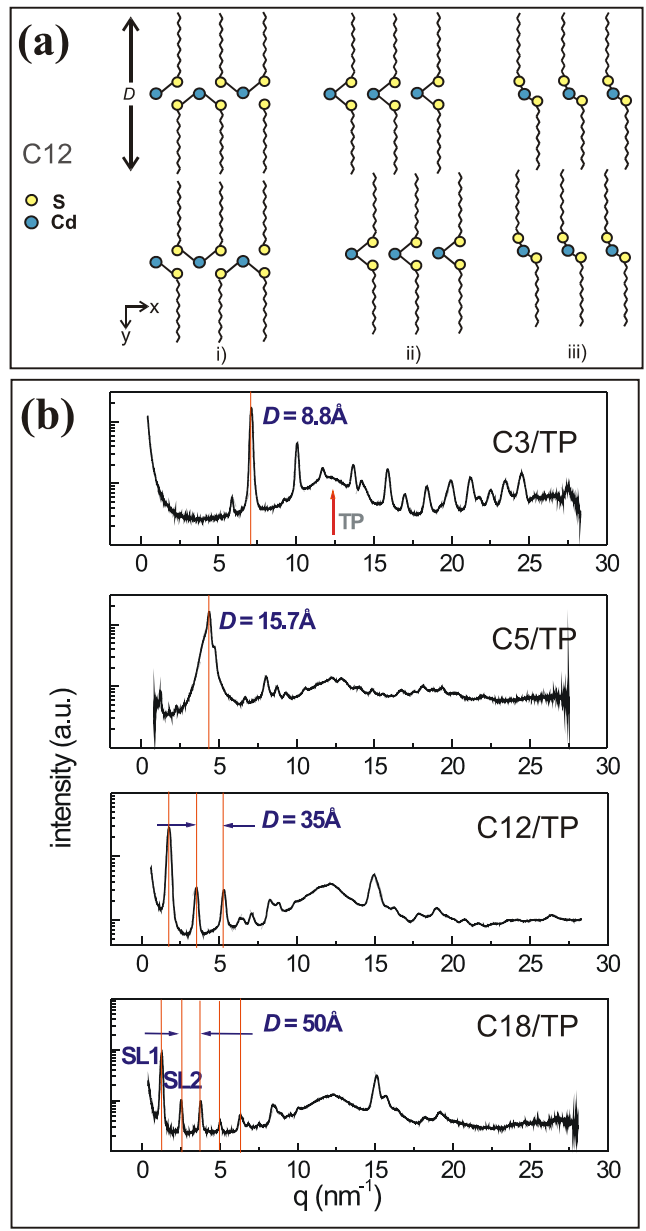

**Figure 2**

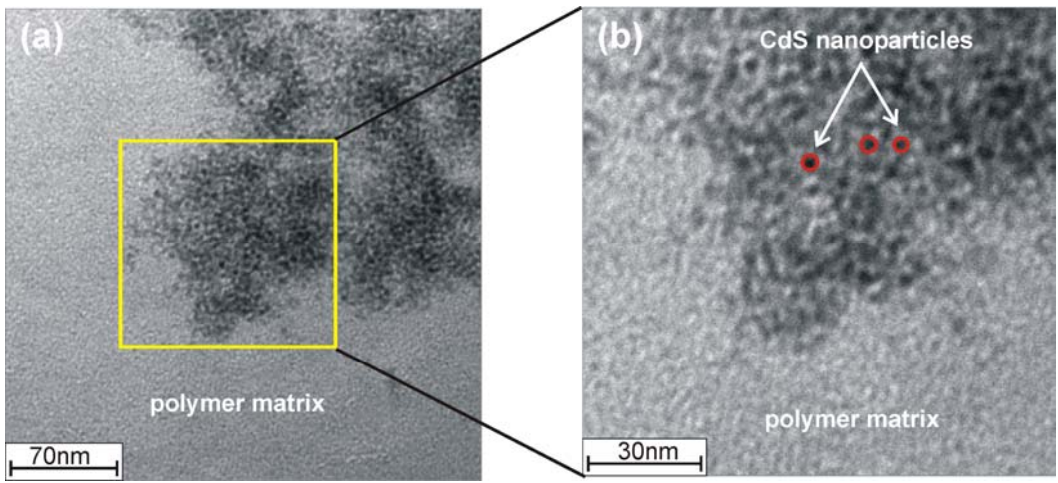

**Figure 3**

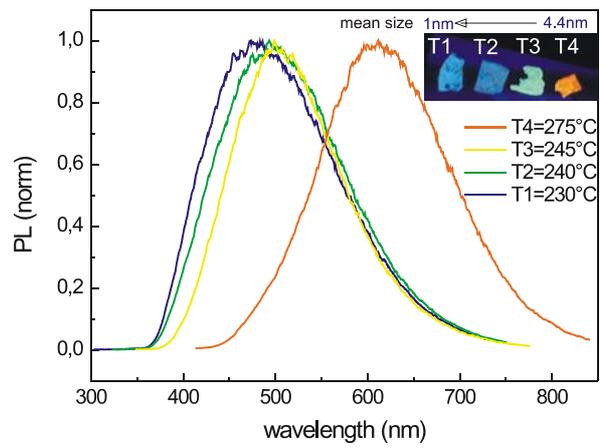

**Figure 4**

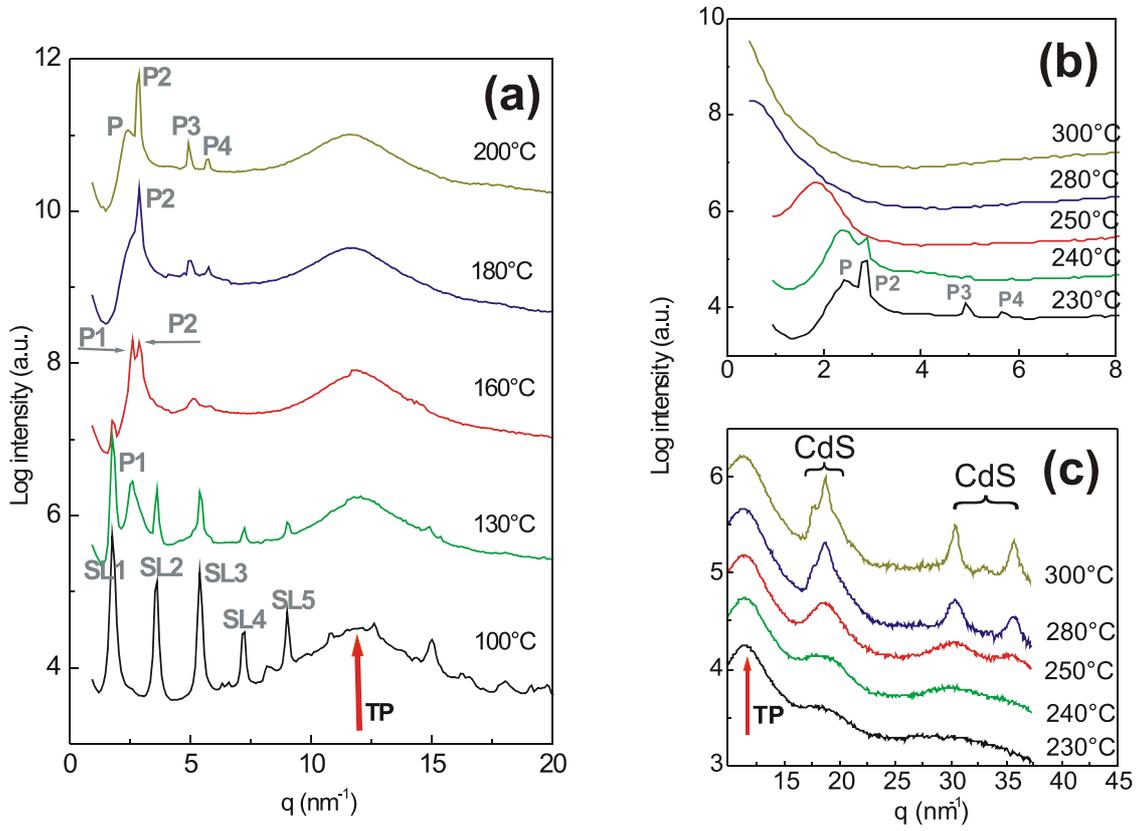

**Figure 5**

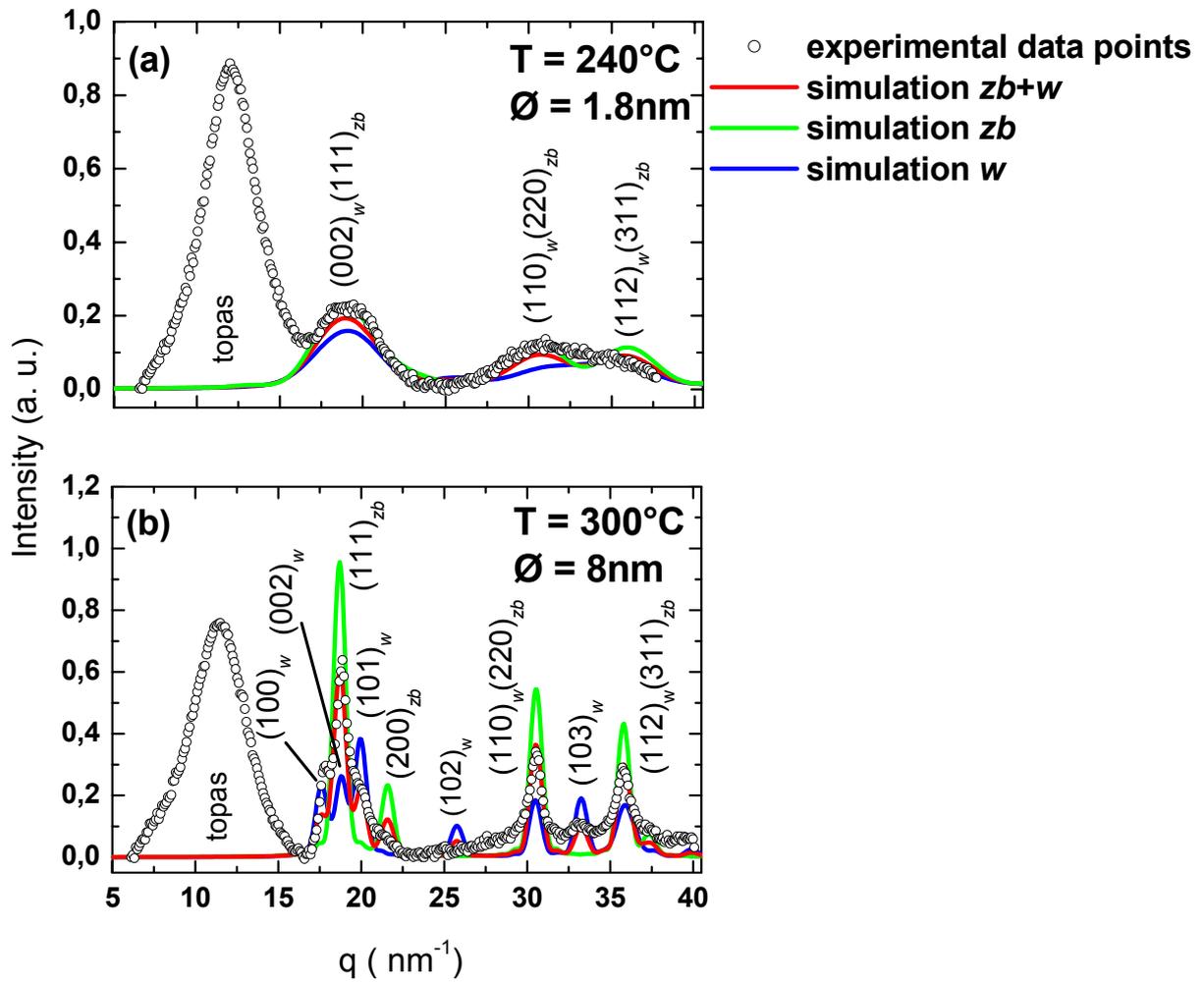

**Figure 6**

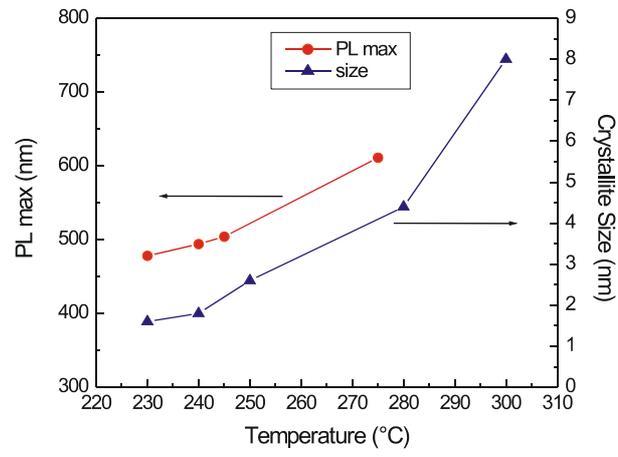

**Figure 7**

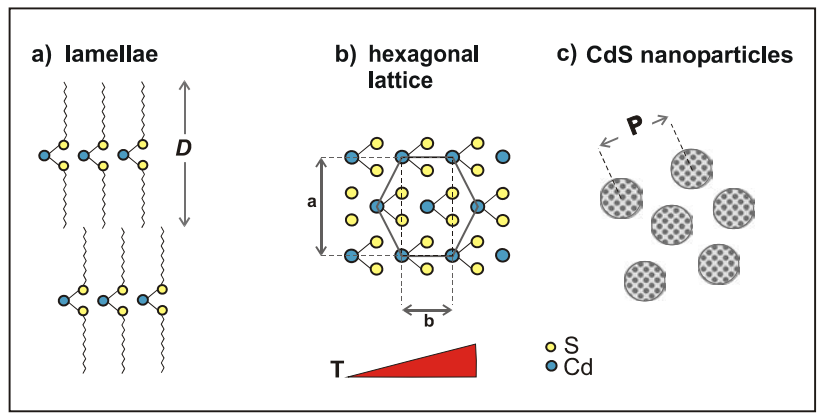